\documentclass[fleqn,10pt]{wlscirep}
\usepackage[utf8]{inputenc}
\usepackage[T1]{fontenc}
\title{Exact path-integral representation of the Wright-Fisher model with mutation and selection}

\author{David Waxman}
\affil{Centre for Computational Systems Biology, ISTBI, Fudan University, 220 Handan Road, Shanghai 200433, China}


\begin{abstract}
The Wright-Fisher model describes a biological population containing a finite
number of individuals. In this work we consider a Wright-Fisher model for a randomly 
mating population, where selection and mutation act at an unlinked locus. The selection 
acting has a general form, and the locus may have two or more alleles.  We determine 
an exact representation of the time dependent transition probability of such a model 
in terms of a \textit{path integral}.  Path integrals were introduced in physics and mathematics, 
and have found numerous applications in different fields, where a probability distribution,
or closely related object, is represented as a `sum' of contributions over all \textit{paths} or
\textit{trajectories} between two points. Path integrals provide alternative
calculational routes to problems, and may be a source of new intuition and suggest 
new approximations. For the case of two alleles, we relate the exact
Wright-Fisher path-integral result to the path-integral form of the transition density 
under the diffusion approximation. We determine properties of the  Wright-Fisher transition probability 
for multiple alleles. We show how, in the absence of mutation, the Wright-Fisher transition 
probability incorporates phenomena such as fixation and loss. 
\end{abstract}
\begin{document}

\flushbottom
\maketitle
\thispagestyle{empty}

\section*{Introduction}

The Wright-Fisher model describes a biological population containing a finite
number of individuals\cite{Fisher1922, Wright}. It represents a fundamental model (or class of models) within
population genetics that continues to be of relevance \cite{EwensBook}. Such
a model, at heart, describes the stochastic fluctuations of allele frequencies
that occur in finite populations. The fluctuations arise because the parents
of one generation do not all make the same contribution to the next
generation. The Wright-Fisher model thus describes random genetic drift.

In addition to random genetic drift, the model can also incorporate the
dynamical effects of selection, mutation and other evolutionary forces, and
has been used in many different situations. For example, the model lies at the
heart of forward simulations\cite{Messer}, as well as in
inference\cite{Tataru}, in evolutionary games\cite{Imhof}, and it connects intimately with the 
coalescent\cite{Fu}.

The Wright-Fisher model is a discrete state, discrete time, Markov chain. The
discrete states correspond to possible allele frequencies (or sets of allele
frequencies), and the discrete time corresponds to generations. The model has
been analysed under the diffusion approximation\cite{Kimura1955,EwensBook}, 
where states and times are treated as taking continuous
values. Recently, the Wright-Fisher model of a biallelic locus, subject to
multiplicative selection but in the absence of mutation, has been considered under
the diffusion approximation, and the time-dependent transition density has been 
represented as a \textit{path-integral}\cite{Schraiber}.

Path integrals were introduced into quantum theory by Feynman\cite{FeynmanHibbs}, 
and into the mathematics of diffusion by Wiener\cite{Kovalchik}. Generally, a probability 
distribution (or closely related object), associated with a diffusion-like process, is
represented as a `sum' of contributions over all \textit{paths} or
\textit{trajectories} between two points or states. The `sum' over paths is
often an integral, and this is the origin of the name `path integral'. An
alternative name is `functional integral', since integration over
paths/trajectories can also be thought of as integration over
\textit{functions} - namely the trajectories themselves. Related approaches,
termed \textit{functional methods}, have a large variety of applications\cite{Kleinert}.

The introduction of path integrals in physics has provided alternative
calculational routes to problems. Beyond this, path integrals, because they focus
on visualisable trajectories, may be a source of new intuition, and may
suggest new ways to proceed and new approximations\cite{FeynmanHibbs}.

In the present paper we work fully within the framework of a Wright-Fisher
model where both states and time are discrete; we do not employ 
the diffusion approximation, although we relate some results to this approximation.
We consider a randomly mating sexual population, where a
general scheme of selection acts at an unlinked locus, which is also
subject to mutation. We derive an exact path-integral representation of the
time-dependent transition probability of this model.

We first consider a locus with two alleles, and then generalise to the locus
having $n$ alleles, where $n$ can take the values $2$, $3$, $4$, ... . There
are numerous examples of studies of loci with two alleles, and there are increasing numbers
of examples where multiple ($>2$) alleles exist at a locus\cite{Okay,Hodge,Campbell,Phillips}.

\section*{Theoretical background for the case of two alleles}

Consider a population that is diploid and sexual, and reproduces by random
mating. We assume there is an equal sex ratio and no sexual dimorphism. Time
is measured in generations, and labelled by $t=0,1,2,3,\ldots$.

The organisms in the population are subject to mutation and selection at a
single unlinked locus. The locus has two alleles that we refer to as $A_{1}$ and
$A_{2}$. We shall focus on just one of the two alleles, say $A_{1}$, and often
call it the \textit{focal allele}.

With two alleles, the sum of their frequencies is unity. Thus specification of the frequency 
of the focal allele at any time determines the frequency of the other allele at this time. The state of the population can thus be described just in terms of the frequency of the focal
allele, which we will often just call the \textit{frequency}.

\subsection*{Effectively infinite population}

We first consider a very large (effectively infinite) population.

We take a generation to begin with adults, who sexually reproduce via random
mating and then die. Each mating yields the same very large number of offspring.

If the frequency of the focal allele in adults is $x$, then the frequency of
the focal allele in offspring is $x^{\ast}=f_{mut}(x)$ where $f_{mut}(x)$
takes into account frequency changes caused by mutation.  This function is given by
\begin{equation}
f_{mut}(x)=\left(  1-u\right)  x+v(1-x) \label{fm}
\end{equation}
where $u$ is the probability that an $A_{1}$ allele in a parent mutates to an
$A_{2}$ allele in an offspring, and $v$ is the corresponding $A_{2}$ to
$A_{1}$ mutation probability. In the absence of mutation $f_{mut}(x)=x$ which
is equivalent to $x^{\ast}=x$.

We assume viability selection determines the probability of different
offspring surviving to maturity. If the frequency of the focal allele in
offspring (i.e., after mutation) is $x^{\ast}$, then the frequency of this
allele after viability selection has acted is $f_{sel}(x^{\ast})$, where
$f_{sel}(x^{\ast})$ takes into account frequency changes of $x^{\ast}$ due to
selection. Some non selective thinning may occur at this point, but providing
the population size remains very large, this does not cause any further
changes in allele frequencies.

Selection acts on variation in the population, and when
there is no variation there are no effects of selection. There is no variation
when carriers of only one allele are present in the population, which
corresponds to $x=0$ and $x=1$. We take
\begin{equation}
f_{sel}(x)=x+\sigma(x)x(1-x) \label{fsel}
\end{equation}
where the function $\sigma(x)$ (with $|\sigma(x)|<\infty$) is determined by
the particular scheme of selection that is operating, and the \textit{effect
of selection} in $f_{sel}(x)$, namely $\sigma(x)x(1-x)$, has the required
property of vanishing at both $x=0$ and $x=1$.

A few examples of $\sigma(x)$ are as follows.

\begin{enumerate}
[label=(\roman*)]

\item If the relative fitnesses of the three genotypes $A_{1}A_{1}$,
$A_{1}A_{2}$ and $A_{2}A_{2}$ are $1+s$, $1+hs$ and $1$, respectively, then
$\sigma(x)=s\times\left[  \left(  1-2h\right)  x+h\right]  /[1+sx^{2}
+2hsx(1-x)]$. To leading order in $s$ (assuming $|s|\ll1$), we have
$\sigma(x)=s\times\left[  \left(  1-2h\right)  x+h\right]  $, in which case
any $h\neq1/2$ will lead to $\sigma(x)$ varying with $x$.

\item If selection is additive, and the relative fitnesses of the three
genotypes are $1+2s$, $1+s$ and $1$, respectively, then 
$\sigma(x)=s/\left(1+2sx\right)$, and to leading order in $s$ we have $\sigma(x)=s$, i.e., a constant.

\item If selection is multiplicative, and the relative fitnesses of the three
genotypes are $\left(  1+s\right)  ^{2}$, $1+s$ and $1$, respectively, then
$\sigma(x)=s/(1+sx)$, and to leading order in $s$ we have $\sigma(x)=s$., i.e., a constant. Thus
weak multiplicative selection is very similar, in effect, to weak additive selection.
\end{enumerate}

We note that while small selection coefficients (i.e., small values of $s$,
and more generally small values of $\sigma(x)$) are common in nature
\cite{EyreWalker}, strongly selected alleles do sometimes occur, for example
alleles that are appreciably deleterious \cite{Okay}. Accordingly, we will
not make the assumption that $|\sigma(x)|$ is small, and will simply assume
that $\sigma(x)$ follows, without any approximation, from a selection scheme.
\bigskip

The frequency of the $A_{1}$ allele in offspring, after selection and mutation
have acted, can be expressed in terms of the frequency, $x$, in adults, as
$f_{sel}(x^{\ast})\equiv f_{sel}(f_{mut}(x))$ and we write
\begin{equation}
F(x)=f_{sel}(f_{mut}(x)). \label{F(x)}
\end{equation}

Let $X(t)$ denote the frequency of the $A_{1}$ allele in the adults of
generation $t$. Then in a very large population, the frequency obeys the
deterministic equation
\begin{equation}
X(t+1)=F(X(t)).
\end{equation}

\subsection*{Finite population}
\label{finite-pop-section}

We now consider a \textit{finite population}, where $N$ adults are present at
the start of each generation. The processes of reproduction, mutation and
viability selection occur as in an effectively infinite population. However,
after viability selection there is a round of non selective sampling/number
regulation of the mature offspring, that leads to $N$ individuals being
present in the population. These become the $N$ adults of the next generation.
The behaviour of this population can be described by a Wright-Fisher model, as is shown in 
textbooks\cite{Hopp}. 
We will now use such a model (which can, like the diffusion approximation, 
incorporate an effective population size\cite{ZhaoGossmannWaxman}).

For the population under consideration, let $\mathbf{M}$ denote the
\textit{transition matrix} of the Wright-Fisher model. We write the $(i,j)$
element of $\mathbf{M}$ as $M_{i,j}$, where $i$ and $j$ can take the values
$0,1,2,\ldots,2N$. Then for a population where the focal allele has the
frequency $j/(2N)$ in one generation, the probability that the focal allele
will have the frequency $i/(2N)$ in the next generation is $M_{i,j}$. With
$\binom{a}{b}=\tfrac{a!}{(a-b)!b!}$ a binomial coefficient, we have\cite{Hopp}
\begin{equation}
M_{i,j}=\binom{2N}{i}\left[  F\left(  \frac{j}{2N}\right)  \right]
^{i}\left[  1-F\left(  \frac{j}{2N}\right)  \right]  ^{2N-i}. \label{Mij}
\end{equation}
The transition matrix is always \textit{normalised}
in the sense that $\sum_{i=0}^{2N}M_{i,j}=1$ for all $j$, and invoking
normalisation can resolve ambiguities (for example, when $F(x)=0$, normalisation
ensures $M_{0,0}=1$ so $\sum_{i=0}^{2N}M_{0,j}=1$ for all $j$). 

The transition matrix is key to many calculations. If $\mathbf{P}(t)$ is a
column vector containing the probabilities of all $2N+1$ possible frequency
states of the population in generation $t$, i.e., the probability distribution
for generation $t$, then using the transition matrix we can determine the
probability distribution for generation $t+1$, namely $\mathbf{P}
(t+1)=\mathbf{MP}(t)$. 
Furthermore, using the elements of the transition
matrix, we can determine the probability that the population passes through a
particular set of frequency states over time, i.e., displays a particular frequency
trajectory. For example, if the population
starts with frequency $l/(2N)$ in one generation, then the probability that in
the next $3$ generations the population will have the frequencies $k/(2N)$,
$j/(2N)$ and $i/(2N)$, respectively, is given by $M_{i,j}M_{j,k}M_{k,l}$.

\subsubsection*{Alternative notation for the transition matrix}

We shall now write elements of $\mathbf{M}$ in a different notation that will
be useful for our purposes. We introduce the notion of an \textit{allowed
frequency} of an allele which is given by
\begin{equation}
\text{allowed frequency}=\frac{\text{integer}}{2N} \label{allowed}
\end{equation}
where the integer can take any of the values $0,1,2,\ldots,2N$.

To keep the notation as simple as possible, we shall, for a locus with two
alleles, reserve the use of $a$, $x$, $x^{\prime}$, $x(r)$ (for various
integral $r$) and $z$, as values of allowed frequencies. In terms of the
allowed frequencies $x^{\prime}$ and $x$, we write the elements of
$\mathbf{M}$ as
\begin{equation}
M(x^{\prime}|x)=\binom{2N}{2Nx^{\prime}}\left[  F(x)\right]  ^{2Nx^{\prime}
}\left[  1-F(x)\right]  ^{2N(1-x^{\prime})} \label{M x' x general}
\end{equation}
which gives the probability of a transition from the population state (i.e.,
frequency) $x$, in one generation, to state $x^{\prime}$ in the next
generation. Thus if $x^{\prime}=i/(2N)$ and $x=j/(2N)$, with $i$ and $j$ any
two of the integers $0,1,2,\ldots,2N$, then $M(x^{\prime}|x)$ coincides with
$M_{i,j}$ in Eq. (\ref{Mij}).

We shall refer to a locus that is \textit{not subject to selection} (but which
may be subject to mutation), as a \textit{neutral locus}. The transition
matrix of a neutral locus, written $M^{(0)}(x^{\prime}|x)$, is obtained from
$M(x^{\prime}|x)$ by setting $\sigma(x)$ to zero for all $x$. With
\begin{equation}
x^{\ast}=f_{mut}(x) \label{x star}
\end{equation}
this leads to
\begin{equation}
M^{(0)}(x^{\prime}|x)=\binom{2N}{2Nx^{\prime}}\left(  x^{\ast}\right)
^{2Nx^{\prime}}\left(  1-x^{\ast}\right)  ^{2N(1-x^{\prime})}. \label{M0}
\end{equation}

A central aspect of the analysis we present is that the form of $F(x)$ in Eq.
(\ref{F(x)}) allows us to write the transition matrix $M(x^{\prime}|x)$ of Eq.
(\ref{M x' x general}) as the exact product of two factors:
\begin{equation}
M(x^{\prime}|x)=M^{(0)}(x^{\prime}|x)\times e^{C(x^{\prime}|x)}
\label{M factorisation}
\end{equation}
where $M^{(0)}(x^{\prime}|x)$ is the neutral result given in Eq. (\ref{M0}),
while, with $x^{\ast}$ given by Eq. (\ref{x star}),  we have
\begin{equation}
C(x^{\prime}|x)=2N\times\big[x^{\prime}\ln\big(1+\sigma(x^{\ast})\left(
1-x^{\ast}\right)  \big)+(1-x^{\prime})\ln\big(1-\sigma(x^{\ast})x^{\ast
}\big)\big] \label{C}
\end{equation}
- see the first subsection of \textbf{Methods} for details.

The factorisation in Eq. (\ref{M factorisation}) says the transition matrix
$M(x^{\prime}|x)$ depends on a `core' neutral/mutational part, $M^{(0)}
(x^{\prime}|x)$, and a factor $e^{C(x^{\prime}|x)}$ that is `selectively
controlled' in the sense that if there is no selection (i.e., if $\sigma(x)$
vanishes for all $x$) then $C(x^{\prime}|x)$ vanishes and the factor
$e^{C(x^{\prime}|x)}$ is simply unity. 

We note that while $C(x^{\prime}|x)$ depends on selection, and precisely vanishes in the 
absence of selection, it also depends on the population size, $N$, and the mutation rates $u$ and $v$.

\subsection*{Trajectories and path integral}

We now consider a \textit{trajectory} of the frequency, which starts at
frequency $x(0)$ in generation $0$, has frequency $x(1)$ in generation
$1$,\ldots, and frequency $x(t)$ in generation $t$. To represent such a
trajectory, which runs from time $0$ to time $t$, we use the notation
\begin{equation}
\lbrack x]_{0}^{t}=\left(  x(0),x(1),...,x(t)\right). \label{n=2 traj}
\end{equation} This expresses the trajectory as a row vector with $t+1$ elements, all of
which are allowed frequencies.

The \textit{probability} of occurrence of the trajectory $[x]_{0}^{t}$ in Eq.
(\ref{n=2 traj}) is obtained by multiplying together the appropriate
$M(x^{\prime}|x)$ and is given by ${\textstyle\prod\limits_{r=1}^{t}
}M(x(r)|x(r-1)={\textstyle\prod\limits_{r=1}^{t}}M^{(0)}(x(r)|x(r-1))\times
e^{\sum_{r=1}^{t}C(x(r)|x(r-1))}$. We write this as
\begin{equation}
\text{probability of }[x]_{0}^{t}=W^{(0)}\left(  [x]_{0}^{t}\right)  \times
e^{C\left(  [x]_{0}^{t}\right)  } \label{W=W0*expC}
\end{equation}
where
\begin{equation}
W^{(0)}\left(  [x]_{0}^{t}\right)  ={\textstyle\prod_{r=1}^{t}}M^{(0)}
(x(r)|x(r-1))
\end{equation}
and
\begin{equation}
C\left(  [x]_{0}^{t}\right)  =\sum_{r=1}^{t}C(x(r)|x(r-1)). \label{C[x]}
\end{equation}

Equation (\ref{W=W0*expC}) says that the trajectory $[x]_{0}^{t}$, in the
presence of selection, has a probability which we can write as the product of
the probability of the trajectory under neutrality, $W^{(0)}\left(
[x]_{0}^{t}\right)  $, and the factor $e^{C\left(  [x]_{0}^{t}\right)  }$
which is selectively controlled. The presence of $e^{C\left(  [x]_{0}
^{t}\right)  }$ in Eq. (\ref{W=W0*expC}) indicates how non-zero selection
(in combination with other forces, via its $N$, $u$ and $v$ dependence),
modifies the probability of occurrence of an entire trajectory under neutrality.

Let $K(z,t|a,0)$ denote the overall probability of going from an initial
allowed frequency of $a$ at time $0$ to the allowed frequency of $z$ at time
$t$. In conventional (Markov chain) language, $K(z,t|a,0)$ is determined from
matrix elements of $\mathbf{M}^{t}$, where $\mathbf{M}$ is the transition
matrix that was introduced above (with elements given in Eq. (\ref{Mij})).
By contrast, in
trajectory language, all possible trajectories, between the end-points ($a$ at
time $0$ and $z$ at time $t$) contribute to $K(z,t|a,0)$. We can thus write
$K(z,t|a,0)$ as sum of the trajectory probability, $W^{(0)}[\mathbf{x}]\times
e^{C\left(  [x]_{0}^{t}\right)  }$, over all possible trajectories. That is
\begin{equation}
K(z,t|a,0)=\overset{x(t)=z}{\underset{x(0)=a}{\sum\cdots\sum}}W^{(0)}\left(
[x]_{0}^{t}\right)  e^{C\left(  [x]_{0}^{t}\right)  }. \label{Exact}
\end{equation}
The notation in Eq. (\ref{Exact}) denotes a sum over all trajectories whose end
points, $x(0)$ and $x(t)$, have the specific allowed frequencies $a$ and $z$,
respectively, while $x(1)$, $x(2)$, $...,x(t-1)$, which give the state of the
population at intermediate times, take values that cover all allowed frequencies.
In Figure 1 we illustrate two trajectories that contribute to a transition probability.

\newpage

\begin{figure}[tbp]
\includegraphics[width=\textwidth]{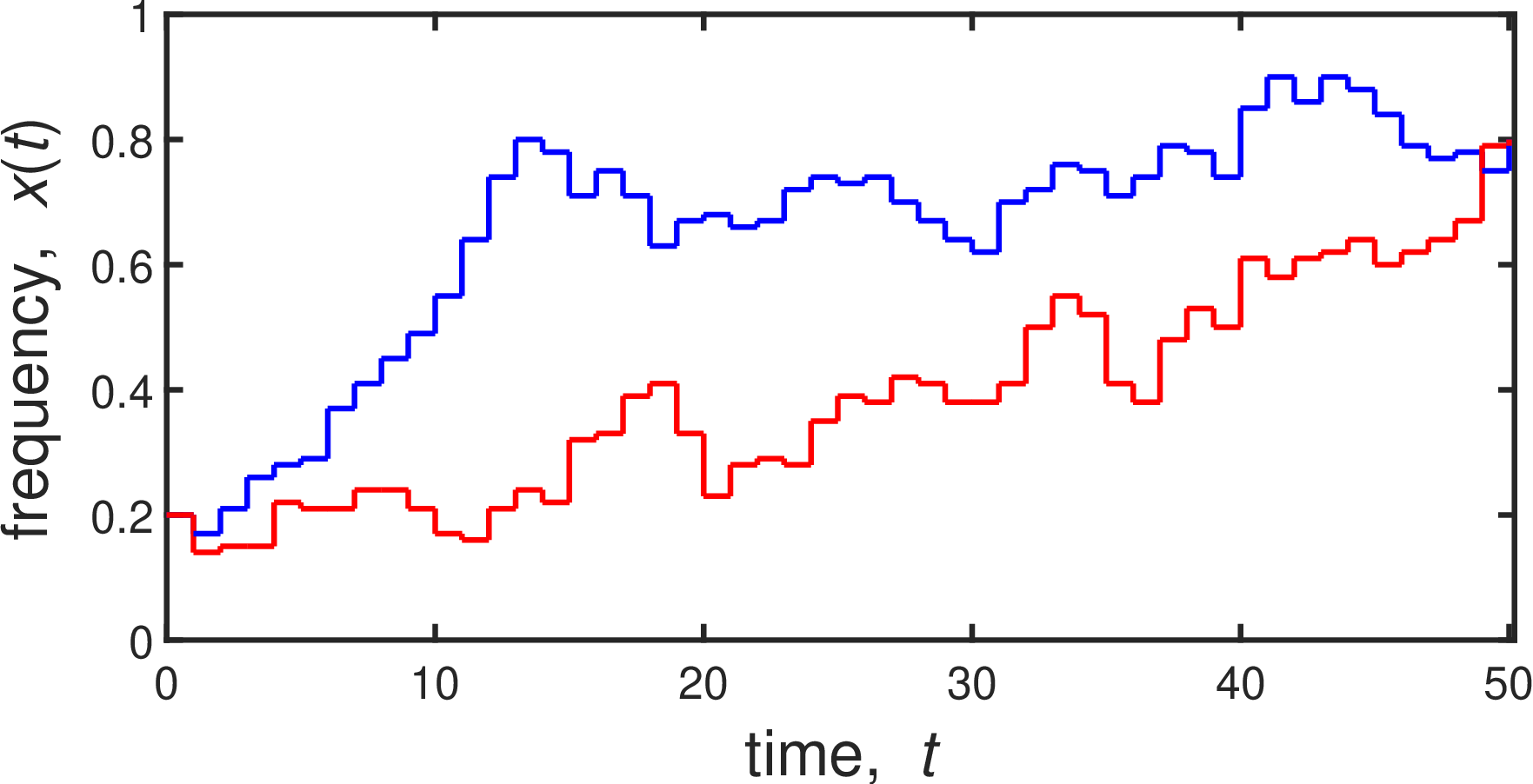} \centering
\caption{\textbf{Contributing trajectories.} An illustration of two \textit{trajectories} (i.e., frequency as 
a function of time) that contribute
to the transition probability $K(z,T|a,0)$, where the initial frequency is $a=0.2$ at time $0$, and the 
final frequency is $z=0.8$ at time $T=50$. All trajectories that contribute to $K(z,T|a,0)$ take only 
\textit{allowed frequencies}.}
\end{figure}

Equation (\ref{Exact}) is an exact `path integral' or `sum over paths'
representation of the finite time transition probability in a two allele
Wright-Fisher model where states and times are discrete, and the model
incorporates mutation and a general form of selection.

Since, by construction, $W^{(0)}\left(  [x]_{0}^{t}\right)  $ is independent
of selection, and since $C\left(  [x]_{0}^{t}\right)  $ vanishes when
selection vanishes, the transition probability corresponding to that in Eq. (\ref{Exact}),
when there is no selection, namely the neutral probability, is written as
$K^{(0)}(z,t|a,0)$ and given by
\begin{equation}
    K^{(0)}(z,t|a,0)=\overset{x(t)=z}{\underset{x(0)=a}
    {\sum\cdots\sum}}W^{(0)}\left([x]_{0}^{t}\right).   \label{K0}
\end{equation}

The quantity $K(z,t|a,0)$ in Eq. (\ref{Exact}) can also be interpreted as the
\textit{probability distribution} of the frequency (of the $A_{1}$ allele) at
time $t$, which is a random variable that we write as $X(t)$. In particular,
$K(z,t|a,0)$ is the value of the distribution of $X(t)$, when evaluated at
frequency $z$, given that the frequency $X(0)$ had the definite value $a$.
Thus, for example, the expected value of $X(t)$, given $X(0)=a$, is
$E[X(t)|X(0)=a]=\sum_{z}z\times K(z,t|a,0)$ where the sum runs over all
allowed values of $z$.

\subsection*{Approximation when there is no mutation and selection is weak}

We now consider a special case of the distribution $K(z,t|a,0)$. We proceed
under the following assumptions.

\begin{enumerate}
[label=(\roman*)]

\item There is no mutation ($u=v=0$).

Equation (\ref{M0}), with no mutation, entails replacing $x^{\ast}$ by $x$ and
we obtain the no mutation, neutral (no selection) form of the transition
matrix that we write as
\begin{equation}
M^{(0,0)}(x^{\prime}|x)=\binom{2N}{2Nx^{\prime}}x^{2Nx^{\prime}}\left(
1-x\right)  ^{2N(1-x^{\prime})}.
\end{equation}

\item Selection is multiplicative.

We take the $A_{1}A_{1}$, $A_{1}A_{2}$ and $A_{2}A_{2}$ genotypes to have
relative fitnesses of $\left(  1+s\right)  ^{2}$, $1+s$ and $1$, respectively.
We then have $\sigma(x)=s/(1+sx)$ and from Eq. (\ref{C}) obtain
\begin{equation}
C(x^{\prime}|x)=2N\left[  x^{\prime}\ln\left(  1+s\right)  -\ln\left(
1+sx\right)  \right]  . \label{mult sel C}
\end{equation}

\item Selection is weak ($|s|\ll1$)

In terms of the \textit{scaled strength of selection}
\begin{equation}
R=2Ns
\end{equation}
which, unlike $s$ need not be small, the expansion of $C(x^{\prime}|x)$ in $s$
is given by $C(x^{\prime}|x)\simeq R\cdot\left(  x^{\prime}-x\right)
-\frac{R^{2}}{4N}\left(  x^{\prime}-x^{2}\right)  $ with corrections of order
$s^{3}$. This yields
\begin{equation}
C\left(  [x]_{0}^{t}\right)\equiv\sum_{r=1}^{t}C(x(r)|x(r-1)
\simeq\left(R-\frac{R^{2}}{4N}\right)\left[  x(t)-x(0)\right]
-\sum_{r=0}^{t-1}U\left(  x(r)\right)                                \label{C2expand}
\end{equation}

where
\begin{equation}
U\left(  x\right)  =\frac{R^{2}}{4N}x\left(  1-x\right)  .
\end{equation}

\end{enumerate}

Thus in the absence of mutation, but with weak selection, we have the
approximation
\begin{equation}
K(z,t|a,0)\simeq e^{\left[  R-R^{2}/(4N)\right]  \left(  z-a\right)
}\overset{x(t)=z}{\underset{x(0)=a}{\sum\cdots\sum}}W^{(0,0)}\left(
[x]_{0}^{t}\right)  e^{-\sum_{r=0}^{t-1}U\left(  x(r)\right)  } .
\label{Approx 2 allele}
\end{equation}

The path integral representation of the
transition probability density, under the \textit{diffusion approximation},
which involves continuous frequencies and continuous time, can be written as 
\begin{equation}
K_{\text{\text{diffusion}}}(z,t|a,0)=e^{R\cdot\left(  z-a\right)  }
\int_{x(0)=a} ^{x(t)=z}P([x]_{0}^{t})e^{-\int_{0}^{t}U(x(r))dr}d[x]
\label{K diff}
\end{equation}
where the integration is over all continuous trajectories that start at
frequency $a$ at time $0$ and end at frequency $z$ at time $t$, with
$P([x]_{0}^{t})$ a `weight' associated with neutral trajectories, and $d[x]$ the measure of the path integral\cite{Schraiber}.

A comparison of the approximate Wright-Fisher transition probability in Eq. (\ref{Approx 2 allele}) 
and the diffusion transition probability density in Eq. (\ref{K diff}) indicates that the two results 
are similar. In particular, corresponding to the expressions $e^{\left[
R-R^{2}/(4N)\right]\left(z-a\right)}$ and $e^{-\sum_{r=1}^{t}U\left(x(r)\right)}$
that are present in the Wright-Fisher result
are, respectively, the expressions $e^{R\cdot\left(z-a\right)}$ and $e^{-\int_{0}^{t}U(x(r))dr}$
in the diffusion result. The analogue of the Wright-Fisher neutral, mutation-free, 
trajectory probability, $W^{(0,0)}\left(  [x]_{0}
^{t}\right)$, that is present in Eq. (\ref{Approx 2 allele}), is the neutral weight, 
$P([x]_{0}^{t})$, that is present in Eq. (\ref{K diff}).

\section*{Theoretical background for multiple alleles}

We shall now generalise the above. We again consider a population that is
diploid, reproduces sexually by random mating, has an equal sex ratio,
exhibits no sexual dimorphism, and evolves in discrete generations. Selection
again occurs at a single unlinked locus, but now there are $n$ alleles at the
locus, where $n$ is arbitrary (i.e., $n=2,3,4,...$) and we write allele $i$
(for $i=1,2,...,n$) as $A_{i}$.

When there are three or more alleles, the difference, compared with two
alleles, is that knowledge of the frequency of one allele is not enough to
specify the state of the population. In fact, we need to follow the behaviour
of $n-1$ allele frequencies, while one allele frequency can be treated as
being determined by all other allele frequencies (since allele frequencies sum
to unity). However, we shall not proceed in this way; we shall treat all
alleles as being on an equal footing, and follow the behaviour of
all $n$ allele frequencies.

\subsection*{Effectively infinite population}

We first consider a very large (effectively infinite) population.

In what follows, we shall use $\mathbf{x}$ to denote an $n$ component column
vector whose $i$'th element, $x_{i}$, is the frequency of allele $A_{i}$ in
adults ($i=1,2,...,n$).

The frequency of all alleles in offspring is then $\mathbf{x}^{\ast}
=\mathbf{Q}\mathbf{x}$ where $\mathbf{Q}$ is an $n\times n$ matrix whose
$(i,j)$ element, $Q_{i,j}$, is the probability that an $A_{j}$ allele mutates
to an $A_{i}$ allele. Elements of $\mathbf{Q}$ are non-negative, and satisfy $\sum_{i=1}^{n}Q_{i,j}=1$ for all $j$
(so the sum of all mutated frequencies is unity).

We next assume that
viability selection acts and determines the probability of different offspring
surviving to maturity. The frequencies, after viability selection, are given
by $\mathbf{f}_{sel}(\mathbf{x}^{\ast})$, where $\mathbf{f}_{sel}
(\mathbf{x}^{\ast})$ takes into account frequency changes of $\mathbf{x}
^{\ast}$ due to selection, and is an $n$ component column vector.

We shall shortly exploit a property of $\mathbf{f}_{sel}(\mathbf{x})$, that
follows because selection acts on variation in a population. In particular, if
the vector of allele frequencies, $\mathbf{x}$, has an $i$'th element which is
zero ($x_{i}=0$), then the $i$'th element of $\mathbf{f}_{sel}(\mathbf{x})$,
which we write, as $f_{sel,i}(\mathbf{x})$, also vanishes, since selection
alone cannot salvage an allele after its absence from a population. This
motivates us to take $f_{sel,i}(\mathbf{x})$ in the form
\begin{equation}
f_{sel,i}(\mathbf{x})=x_{i}\times [1+G_{i}(\mathbf{x})] \label{fsel,i=xi*Gi}
\end{equation}
where $G_{i}(\mathbf{x})$ is finite ($\left\vert G_{i}(\mathbf{x})\right\vert
<\infty$) and is determined by the specific form
of selection acting. Generally, $\sum_{i=1}^{n}x_{i}G_{i}(\mathbf{x})=0$ and $G_{i}(\mathbf{x})\geq-1$
(ensuring that after selection, the sum of all allele frequencies is unity, and all 
alleles frequencies are non-negative).

The set of allele frequencies in offspring, after selection and mutation have
acted, can be expressed in terms of the set of frequencies $\mathbf{x}$, in
adults, as $\mathbf{f}_{sel}(\mathbf{Qx})$ and we write
\begin{equation}
\mathbf{F}(\mathbf{x})=\mathbf{f}_{sel}(\mathbf{Qx}). \label{F multi}
\end{equation}

We now consider dynamics. Let $\mathbf{X}(t)$ denote an $n$ component column
vector containing the set of allele frequencies in generation $t$. Because we
have an effectively infinite population, $\mathbf{X}(t)$ obeys the
deterministic equation
\begin{equation}
\mathbf{X}(t+1)=\mathbf{F}(\mathbf{X}(t)). \label{multiallele deterministic}
\end{equation}

\subsection*{Finite population}
Consider now a finite population, where $N$ adults are present in each
generation. The quantity $\mathbf{x}$ is still an $n$ component vector whose
$i$'th element, $x_{i}$, is the frequency of allele $A_{i}$ in adults, but it
has the added feature that all elements have values which are allowed
frequencies (Eq. (\ref{allowed})). That is, $x_{i}\geq0$, $\sum_{i=1}^{n}
x_{i}=1$, and each $x_{i}$ is an integer divided by $2N$. We shall call a
vector that has this property an \textit{allowed set} of allele frequencies.
In the multiallele case we shall reserve the use of $\mathbf{a}$, $\mathbf{x}
$, $\mathbf{x}^{\prime}$, $\mathbf{x}(r)$ (for various $r$), and $\mathbf{z}$,
for \textit{allowed sets} of allele frequencies. We now write the transition
matrix element for the probability of a transition from state $\mathbf{x}$ to
state $\mathbf{x}^{\prime}$ as
\begin{equation}
M(\mathbf{x}^{\prime}|\mathbf{x})=\binom{2N}{2N\mathbf{x}^{\prime}}
\prod\limits_{i=1}^{n}\left[  F_{i}(\mathbf{x})\right]  ^{2Nx_{i}^{\prime} }
\label{M multi}
\end{equation}
where $\binom{2N}{\mathbf{m}}$, with $\mathbf{m}$ an $n$ component column vector with
integer elements, denotes a multinomial coefficient for $n$
categories. We note that the transition matrix element, $M(\mathbf{x}^{\prime}|\mathbf{x})$, in its
conventional matrix form, is an element of a matrix with \textit{vector} indices, not scalars\cite{Waxman2009}.

The transition matrix of a neutral locus has elements which are the zero
selection limit of $M(\mathbf{x}^{\prime}|\mathbf{x})$, which we write as
$M^{(0)}(\mathbf{x}^{\prime}|\mathbf{x})$, and which is given by
\begin{equation}
M^{(0)}(\mathbf{x}^{\prime}|\mathbf{x})=\binom{2N}{2N\mathbf{x}^{\prime}}
\prod\limits_{i=1}^{n}\left(  x_{i}^{\ast}\right)  ^{2Nx_{i}^{\prime} }
\label{M0 multi}
\end{equation}
where $\mathbf{x}^{\ast}$ is given by
\begin{equation}
\mathbf{x}^{\ast}=\mathbf{Qx}. \label{x* multi}
\end{equation}

As for the case of two alleles, a factorisation is possible; the form of
$\mathbf{f}_{sel}(\mathbf{x})$ in Eq. (\ref{fsel,i=xi*Gi}) allows us to write
the transition matrix, $M(\mathbf{x}^{\prime}|\mathbf{x})$ of Eq.
(\ref{M multi}), as the exact product of two factors:
\begin{equation}
M(\mathbf{x}^{\prime}|\mathbf{x})=M^{(0)}(\mathbf{x}^{\prime}|\mathbf{x}
)\times e^{C(\mathbf{x}^{\prime}|\mathbf{x})} \label{M multi factorisation}
\end{equation}
where $M^{(0)}(\mathbf{x}^{\prime}|\mathbf{x})$ is given in Eq.
(\ref{M0 multi}) and
\begin{equation}
C(\mathbf{x}^{\prime}|\mathbf{x})=2N\sum_{i=1}^{n}x_{i}^{\prime}
\ln\big(1+G_{i}(\mathbf{x}^{\ast})\big) \label{C x' x}
\end{equation}
- see the second subsection of \textbf{Methods} for details.

\subsection*{Trajectories and path integral}

We now write a trajectory as
\begin{equation}
\lbrack\mathbf{x}]_{0}^{t}=\left(  \mathbf{x}(0),\mathbf{x}(1),...,\mathbf{x}
(t)\right)  \label{general n trajectory}
\end{equation}
in which each $\mathbf{x}(r)$ is an $n$ component column vector containing an
\textit{allowed set} of allele frequencies, which gives the state of the
population at time $r$. It follows that the trajectory $[\mathbf{x}]_{0}^{t}$
in Eq. (\ref{general n trajectory}), is an $n\times(t+1)$ matrix. The
probability of this trajectory is ${\textstyle\prod\limits_{r=1}^{t}}
M(\mathbf{x}(r)|\mathbf{x}(r-1)= {\textstyle\prod\limits_{r=1}^{t}}
M^{(0)}(\mathbf{x}(r)|\mathbf{x}(r-1))\times\exp\left(  \sum_{r=1}
^{t}C(\mathbf{x}(r)|\mathbf{x}(r-1))\right)  $. We write this as
\begin{equation}
\text{probability of}\ [\mathbf{x}]_{0}^{t}=W^{(0)}\left(  [\mathbf{x}
]_{0}^{t}\right)  \times e^{C\left(  [\mathbf{x}]_{0}^{t}\right)  }
\label{prob X}
\end{equation}
where
\begin{equation}
W^{(0)}\left(  [\mathbf{x}]_{0}^{t}\right)  = {\textstyle\prod_{r=1}^{t}}
M^{(0)}(\mathbf{x}(r)|\mathbf{x}(r-1)) \label{W0 multi}
\end{equation}
and
\begin{equation}
C\left(  [\mathbf{x}]_{0}^{t}\right)  =\sum_{r=1}^{t}C(\mathbf{x}
(r)|\mathbf{x}(r-1)).   \label{C multi}
\end{equation}

Let $K(\mathbf{z},t|\mathbf{a},0)$ denote the overall probability of going
from an initial state of the population corresponding to the allowed set of
frequencies, $\mathbf{a}$ at time $0$, to state $\mathbf{z}$ at time $t$,
which is an another allowed set of frequencies. All possible trajectories
between these end-points contribute to $K(\mathbf{z},t|\mathbf{a},0)$. We thus
write $K(\mathbf{z},t|\mathbf{a},0)$ as sum of the probabilities
$W^{(0)}\left(  [\mathbf{x}]_{0}^{t}\right)  \times e^{C\left(  [\mathbf{x}
]_{0}^{t}\right)  }$ over all possible trajectories. That is
\begin{equation}
K(\mathbf{z},t|\mathbf{a},0)=\overset{\mathbf{x}(t)=\mathbf{z}
}{\underset{\mathbf{x}(0)=\mathbf{a}}{\sum\cdots\sum}}W^{(0)}\left(
[\mathbf{x}]_{0}^{t}\right)  e^{C\left(  [\mathbf{x}]_{0}^{t}\right)  }.
\label{Exact multallele}
\end{equation}
The notation in Eq. (\ref{Exact multallele}) denotes a sum over all trajectories
whose end points, $\mathbf{x}(0)$ and $\mathbf{x}(t)$, have the specific
(\textit{allowed set}) values $\mathbf{a}$ and $\mathbf{z}$, respectively,
while $\mathbf{x}(1)$, $\mathbf{x}(2)$, $...,\mathbf{x}(t-1)$, which give the
state of the population at intermediate times, take values that cover all
\textit{allowed sets }of frequencies.

Equation (\ref{Exact multallele}) is an exact `path integral' representation
of the finite time transition probability in a multiple ($n$) allele
Wright-Fisher model where states and times are discrete.

Since, by construction, $W^{(0)}\left(  [\mathbf{x}]_{0}^{t}\right)  $ is
independent of selection, and since $C\left(  [\mathbf{x}]_{0}^{t}\right)  $
vanishes when there is no selection, the probability of going from state
$\mathbf{a}$ at time $0$ to state $\mathbf{z}$ at time $t$, when there is no
selection, is $K^{(0)}(\mathbf{z},t|\mathbf{a},0)=\overset{\mathbf{x}
(t)=\mathbf{z}}{\underset{\mathbf{x}(0)=\mathbf{a}}{\sum\cdots\sum}}
W^{(0)}\left(  [\mathbf{x}]_{0}^{t}\right)  $.

\subsection*{Approximation when there is no mutation and selection is weak}

We now consider a special case of the distribution $K(\mathbf{z}
,t|\mathbf{a},0)$ of Eq. (\ref{Exact multallele}), when there is no mutation
and selection is multiplicative and weak.

When there is no mutation the matrix $\mathbf{Q}$ becomes the $n\times n$
identity matrix.

Under multiplicative selection, we take the $A_{i}A_{j}$ genotype to have a
fitness proportional to $(1+s_{i})(1+s_{j})$. It follows that $F_{i}
(\mathbf{x})=x_{i}\left(  s_{i}-{\sum\nolimits_{j=1}^{n}}s_{j}x_{j}\right)
/\left(  1+{\sum\nolimits_{j=1}^{n}}s_{j}x_{j}\right)  $ hence $G_{i}
(\mathbf{x})=\left(  s_{i}-{\sum\nolimits_{j=1}^{n}}s_{j}x_{j}\right)
/\left[  1+{\sum\nolimits_{j=1}^{n}}s_{j}x_{j}\right]  $ and
\begin{equation}
C(\mathbf{x}^{\prime}|\mathbf{x})=2N\left[  \sum_{i=1}^{n}x_{i}^{\prime}
\ln\left(  1+s_{i}\right)  -\ln\left(  1+\sum_{i=1}^{n}s_{i}x_{i}\right)
\right]  .
\end{equation}

We take weak selection to correspond to $|s_{i}|\ll1$ for all $i$, then
similar to the case of two alleles, we obtain approximate results by expanding
$C(\mathbf{x}^{\prime}|\mathbf{x})$ in the $s_{i}$, and discarding third and
higher order terms. We shall express results in terms of scaled selection
strengths that are given by
\begin{equation}
\mathbf{R}=2N\mathbf{s}\text{ or }R_{i}=2Ns_{i} \label{R multi} 
\end{equation} 
where $\mathbf{s}$ is a column vector of the $s_{i}$.

With $\delta_{i,j}$ denoting a Kronecker delta, a $T$ superscript denoting the
transpose of a vector, and $\mathbf{V}(\mathbf{x})$ denoting an $n\times n$
matrix with elements
\begin{equation}
V_{i,j}(\mathbf{x})=x_{i}\delta_{i,j}-x_{i}x_{j}
\end{equation}
we obtain
\begin{equation}
C\left(  [\mathbf{x}]_{0}^{t}\right)  \equiv\sum_{r=1}^{t}C(\mathbf{x}
(r)|\mathbf{x}(r-1)
\simeq\mathbf{R}^{T}\left[  \mathbf{x}(t)-\mathbf{x}(0)\right]
+\phi\left(  \mathbf{x}(t)\right)  -\phi\left(  \mathbf{x}(0)\right)
-\sum_{r=0} ^{t-1}U\left(  \mathbf{x}(r)\right)  \label{C  mult}
\end{equation}
where
\begin{equation}
\phi\left(  \mathbf{x}\right)  =-\sum_{i=1}^{n}\frac{R_{i}^{2}}{4N} x_{i}
\label{phi}
\end{equation}
and
\begin{equation}
U\left(  \mathbf{x}\right)  =\frac{1}{4N}\mathbf{R}^{T}\mathbf{V}
(\mathbf{x})\mathbf{R}. \label{U mult}
\end{equation}

Using Eqs. (\ref{C mult}), (\ref{phi}) and (\ref{U mult}) in Eq.
(\ref{Exact multallele}), combined with $W^{(0,0)}\left(  [\mathbf{x}]_{0}
^{t}\right)$, which denotes the probability of trajectory $[\mathbf{x}
]_{0}^{t}$ in the absence of mutation and selection ($W^{(0,0)}\left([\mathbf{x}]_{0}^{t}\right)$ 
is constructed from a product of terms of the form $\binom{2N}{2N\mathbf{x}^{\prime}} \prod\limits_{i=1}
^{n}x_{i}^{2Nx_{i}^{\prime} }$ - cf. Eq. (\ref{W0 multi})).
we obtain the
approximation
\begin{equation}
K(\mathbf{z},t|\mathbf{a},0)\simeq e^{\mathbf{R}^{T}\left(  \mathbf{z}
-\mathbf{a}\right)  +\phi\left(  \mathbf{z}\right)  -\phi\left(
\mathbf{a}\right)  }\overset{\mathbf{x}(t)=\mathbf{z}}{\underset{\mathbf{x}
(0)=\mathbf{a}}{\sum\cdots\sum}}W^{(0,0)}\left(  [\mathbf{x}]_{0}^{t}\right)
e^{-\sum_{r=0}^{t-1}U\left(  \mathbf{x}(r)\right)  }.
\label{Approx multiallele}
\end{equation}

\section*{Discussion}

In this work we have derived an exact `path integral' representation of the
time-dependent transition probability in a Wright-Fisher model. We have
explicitly considered the case of two alleles, where the population's
description is in terms of a focal allele, and the case of an arbitrary number
of $n$ alleles, where the description is in terms of all $n$ allele
frequencies, with all frequencies treated as having the same status.

For the case of two alleles, we have compared the Wright-Fisher transition
probability with a path integral representation of the corresponding quantity
(a transition density) under the diffusion approximation. The result for the
diffusion approximation result was derived for multiplicative selection, in
the absence of mutation, and we have established the relation of this with the
exact Wright-Fisher result in this case.

The Wright-Fisher path integral, derived in this work for two alleles, applies
for a wider class of fitness functions than just multiplicative fitness, and
can incorporate mutation. The general form of the path integral, for two
alleles is given in Eq. (\ref{Exact}), and takes the form of a sum over
trajectories of a product the two terms: (i) a `weight' $W^{(0)}\left(
[\mathbf{x}]_{0}^{t}\right)  $ which gives the probability of a trajectory
under neutrality, i.e., when only random genetic drift and mutation are
operating, and (ii) the factor $e^{C\left(  [\mathbf{x}]_{0}^{t}\right)  }$
which while depending on parameters such as mutation rates, is primarily
determined by selection - this factor incorporates all effects of selection,
and $C\left(  [\mathbf{x}]_{0}^{t}\right)  $ vanishes in the absence of
selection. This separation into two factors represents an underlying property
of the transition probability, $K(z,t|a,0)$, that we know from other analyses,
namely that at \textit{long times} ($t\rightarrow\infty$) the quantity
$K(z,t|a,0)$ is a smooth function of selection, but the long time properties
are very different for zero and non-zero mutation rates. For non-zero mutation
rates, the long-time form of $K(z,t|a,0)$ is non-zero for all possible values
of $z$ (i.e., all allowed frequencies), and independent of the initial
frequency, $a$. By contrast, for vanishing mutation rates, only the terminal
frequency classes ($0$ and $1$) have non-zero probabilities at long times, and
furthermore, these probabilities depend on the initial frequency, $a$. Thus
$K(z,t|a,0)$, as $t\rightarrow\infty$, behaves discontinuously, as a function
of mutation rates, in the sense that allowing mutation rates to tend to zero, 
and having mutation rates exactly equal to zero, yield different results. 
A diffusion analysis shows this most clearly, where
singular spikes (Dirac delta functions) at the terminal frequencies are
generally present in the transition probability density when mutation rates
are zero, and are absent when mutation rates are non-zero \cite{McKane_Waxman}. 
The separation of a probability of a
trajectory into the product of $W^{(0)}\left(  [\mathbf{x}]_{0}^{t}\right)  $
and $e^{C\left(  [\mathbf{x}]_{0}^{t}\right)  }$ is thus natural and a reflection
of different behaviours arising from different features of the dynamics.

On the matters of fixation and loss, we note that since a Wright-Fisher model can
describe these phenomena (in the absence of mutation), an exact path
integral representation associated with this model can also, generally,
describe features such as fixation and loss. This will also carry over to a
path integral representation, based on the diffusion approximation, since the
diffusion approximation is also known to encompass fixation and loss, albeit
in a singular form \cite{McKane_Waxman}. Such singular behaviour seems likely
to make the analysis of the path integral representation, based on the diffusion approximation, 
to be more complex, than in its absence.

As an elementary illustration of how fixation is incorporated into the path
integral representation of the transition probability, $K(z,t|a,0)$, we note
the when all mutation rates are zero, the probability of ultimate fixation of
the focal allele is $\lim_{t\rightarrow\infty}K(1,t|a,0)$. Let us revisit the
case
considered above
where there is no mutation and
weak multiplicative selection acting. We can expand $K(1,t|a,0)$ in $s$ by
first expanding $K(1,t|a,0)$ in $C\left(  \left[  x_{0}^{t}\right]  \right)
$, and then expanding $C\left(  \left[  x_{0}^{t}\right]  \right)$ in $s$.
To linear order in $s$ we obtain (from Eq. (\ref{Approx 2 allele}))
$K(1,t|a,0)\simeq[1+2Ns(1-a)] \times\overset{x(t)=1}{\underset{x(0)=a}{\sum
\cdots\sum}}W^{(0,0)}\left(  [x]_{0}^{t}\right)$.
Since $\lim_{t\rightarrow\infty}\overset{x(t)=1}{\underset{x(0)=a}{\sum
\cdots\sum}}W^{(0,0)}$ is the probability of fixation ultimately occurring,
under neutrality, this limit thus coincides with the initial frequency, $a$.
In this way, we arrive at a fixation probability of
$P_{fix}(a)\simeq a+2Nsa(1-a)$, which contains the neutral result and a term which is
first order in $s$, which is the leading correction due to selection. Expansion of $K(z,t|a,0)$ (and related
quantities) to higher order in $s$, can be achieved, again by exploiting the
factorisation between drift/mutation and selection that occurs in Eq.
(\ref{Exact}). Expansions in $s$ beyond linear order involve more complicated
calculations than that of the linear case.

In the case of two alleles, we have seen the relation between the path
integral of the `fully discrete' Wright-Fisher model and the path integral of
the diffusion approximation, for this model. For the case of an arbitrary number
of $n$ alleles there is, at the present time, no such path integral for the
diffusion approximation. However, from the lessons learned for two alleles we
can infer this some of the properties of the general $n$ case, under the
diffusion approximation. In particular, when selection is multiplicative, and
in the absence of mutation. we infer from Eq. (\ref{Approx multiallele}) that
\begin{equation}
K_{\text{diffusion}}(\mathbf{z},t|\mathbf{a},0) = e^{\mathbf{R}^{T}\left(
\mathbf{z} -\mathbf{a}\right)  } \int_{\mathbf{x}(0)=\mathbf{a}}
^{\mathbf{x}(t)=\mathbf{z}} P([\mathbf{x}]_{0}^{t})e^{-\int_{0}^{t}U\left(
\mathbf{x}(r)\right)  dr}d[\mathbf{x}] \label{diff multiallele}
\end{equation}
where $\mathbf{R}$ is a column vector containing the set of scaled selection
strengths (Eq. (\ref{R multi})), the quantity $P([\mathbf{x}]_{0}^{t})$ is the
analogue of the neutral, mutation-free, probability of a trajectory in a
Wright-Fisher model, $W^{(0,0)}\left(  [\mathbf{x}]_{0}^{t}\right)$, while
$U\left(\mathbf{x}\right)$ is given by Eq. (\ref{U mult}). An interesting
feature is the way selection enters Eq. (\ref{diff multiallele}), in both the
prefactor, $e^{\mathbf{R}^{T}\left(  \mathbf{z}-\mathbf{a}\right)  }$ and
within $U\left(  \mathbf{x}\right)$ in forms that involve the vectors and matrices that occur
in the problem. Additionally, a diffusion analysis would suggest
that all occurrences of the population size, $N$, are replaced by the
effective population size, $N_{e}$.  


In the special cases considered above, of no mutation and weak selection, the `selectively controlled' 
quantities $C\left([x]_{0}^{t}\right)$  and $C\left(  [\mathbf{x}]_{0}^{t}\right)$, for two and $n$ alleles, 
respectively, both naturally split into two terms (see Eqs. (\ref{C2expand}) and (\ref{C  mult})). 
One of the terms has dependence on only the initial and final frequencies of the transition probability, and
has no dependence of the frequencies taken by trajectories at intermediate times; it is natural to call this a \textit{
boundary term}. To leading order in selection coefficients, the boundary term changes sign when the sign of
all selection coefficients are \textit{reversed} (for two alleles reversal entails $s\rightarrow-s$; 
for $n$ alleles, $\mathbf{s}\rightarrow-\mathbf{s}$). The boundary term is thus the primary place that the 
deleterious or beneficial effect of a mutation manifests itself. The other term ($U\left([x]_{0}^{t}\right)$ and
$U\left([\mathbf{x}]_{0}^{t}\right)$, respectively)
depends on the frequencies taken by trajectories at \textit{all} times, from 
the initial time to the final time. The $U$ terms, when large, have the
effect of suppressing the contribution of a trajectory. They are a
manifestation of the `probabilistic cost of selection' of an entire
trajectory. Interestingly, the $U$ terms cannot take negative values and
remain unaltered when the sign of all selection coefficients are reversed. 

In summary, we have presented an exact representation of the transition
probability of a Wright-Fisher model in terms of a path integral (in reality a
\textit{sum} over paths/trajectories). Let us conclude with some possible ways
that the path integral representation may be of use. We shall restrict our
considerations to the case of two alleles, where the main result is given in
Eq. (\ref{Exact}), since very similar considerations apply to the $n$ allele
case in Eq. (\ref{Exact multallele}).

\begin{enumerate}
\item The path integral representation may make it easy to carry out an 
expansion in a small parameter, such as a selection coefficient. 
This has been carried out for the transition density at intermediate frequencies, 
under the diffusion approximation\cite{Schraiber}. 
In the present work we have shown that expansion in selection coefficients can also be applied to 
phenomena such as fixation and loss. There may be many other applications of expansion in a 
small parameter..

\item A path integral involves trajectories whose contributions generally have
different probabilities of occurrence. A possible approximation is where the
most probable trajectory, along with trajectories that have with small fluctuations around the most
probable trajectory, are used to estimate the path integral.  The most probable trajectory may
be of interest in its own right, since it may typify the way the population makes a transition between 
two states of the population over time. 

\item The path integral representation involves a fundamental separation of mutation and drift from the 
process of what is primarily selection, as manifested by the two factors $W^{(0)}\left(  [x]_{0}^{t}\right)$ 
and $e^{C\left([x]_{0}^{t}\right)}$ in Eq. (\ref{Exact}). To exploit this separation, we
note that while, in this work, we have implicitly assumed that all parameters 
are independent of the time, a straightforward generalisation of the exact results allows 
parameters to be time dependent. Then one immediate case of application occurs when just selection fluctuates over 
time, with selection coefficients drawn each generation from a given distribution, or generated by a random process.  
In the absence of further knowledge, it is plausibly the case that the 
relevant transition probability follows from an \textit{average} over all such selection coefficients. 
With the average denoted by an overbar, the average of Eq. (\ref{Exact}) reads $\overline{K(z,t|a,0)}
=\overset{x(t)=z}{\underset{x(0)=a}{\sum\cdots\sum}}W^{(0)}\left(  [x]_{0}^{t}\right) \overline{e^{C\left(  
[x]_{0}^{t}\right)  }}$. Thus only the selectively controlled factor is averaged, and this may lead to an 
effective theory that has new/modified selective terms, compared with the case where selection coefficients 
are simply set equal to the time-averaged value\cite{Takahata,Waxman}.

\item 
A different approach, compared to the above three approaches, is to rewrite
Eq. (\ref{Exact}) in the form $K(z,t|a,0)=K^{(0)}(z,t|a,0)\times D(z,t|a,0)$
where $K^{(0)}(z,t|a,0)$ is the neutral result (Eq. (\ref{K0})) and
\begin{equation}
D(z,t|a,0)=\left.  \overset{x(t)=z}{\underset{x(0)=a}{\sum\cdots\sum}}%
W^{(0)}\left(  [x]_{0}^{t}\right)  e^{C\left(  [x]_{0}^{t}\right)  }\right/
\overset{x(t)=z}{\underset{x(0)=a}{\sum\cdots\sum}}W^{(0)}\left(  [x]_{0}%
^{t}\right)  .\label{D}%
\end{equation}
We can interpret $D(z,t|a,0)$ as an average of the quantity $e^{C\left(
[x]_{0}^{t}\right)  }$ over all neutral trajectories that start at allowed
frequency $a$ at time $0$ and end at allowed frequency $z$ at time $t$.
Applying Jensen's inequality\cite{Ross} to Eq. (\ref{D}) yields
$D(z,t|a,0)\geq D_{J}(z,t|a,0)$ with
\begin{equation}
D_{J}(z,t|a,0)=\exp\left(  \left.  \overset{x(t)=z}{\underset{x(0)=a}{\sum
\cdots\sum}}W^{(0)}\left(  [x]_{0}^{t}\right)  C\left(  [x]_{0}^{t}\right)
\right/  \overset{x(t)=z}{\underset{x(0)=a}{\sum\cdots\sum}}W^{(0)}\left(
[x]_{0}^{t}\right)  \right)  .
\end{equation}
Thus we find $K(z,t|a,0)\geq K^{(0)}(z,t|a,0)\times D_{J}(z,t|a,0)$, where the
exponent of $D_{J}(z,t|a,0)$ is a conditional average of $C\left(  [x]_{0}%
^{t}\right)  $ over all \textit{neutral trajectories} that start at frequency
$a$ at time $0$ and end at frequency $z$ at time $t$
\end{enumerate}

\section*{Methods}

Here we give details of the calculations underlying Eqs. (\ref{M factorisation})
and (\ref{M multi factorisation}).

\subsection*{Factorisation of the transition matrix: two alleles}

For the case of two alleles, the transition matrix can be expressed as a 
product of two factors, one of which is independent of selection.

We begin with Eq. (\ref{M x' x general}) for the transition matrix, which we reproduce here for
convenience. We have
\begin{equation}
M(x^{\prime}|x)=\binom{2N}{2Nx^{\prime}}\left[  F(x)\right]  ^{2Nx^{\prime}
}\left[  1-F(x)\right]  ^{2N(1-x^{\prime})} \label{M x' x general appendix}
\end{equation}
where
\begin{equation}
F(x)=f_{sel}(f_{mut}(x)).
\end{equation}
In the absence of selection, $F(x)$ reduces to $f_{mut}(x)$ and $M(x^{\prime
}|x)$ reduces $M^{(0)}(x^{\prime}|x)$, as given by
\begin{equation}
M^{(0)}(x^{\prime}|x) =\binom{2N}{2Nx^{\prime}}\left[  f_{mut}(x)\right]
^{2Nx^{\prime}}\left[  1-f_{mut}(x)\right]  ^{2N(1-x^{\prime})}
\equiv\binom{2N}{2Nx^{\prime}}\left(  x^{\ast}\right)  ^{2Nx^{\prime}
}\left(  1-x^{\ast}\right)  ^{2N(1-x^{\prime})}
\end{equation}
where we have set
\begin{equation}
x^{\ast}=f_{mut}(x).
\end{equation}

To establish factorisation we use the adopted form of selection in Eq.
(\ref{fsel}), namely $f_{sel}(x)=x+\sigma(x)x(1-x)$ to write
$F(x)=x^{\ast}\left[  1+\sigma(x^{\ast})(1-x^{\ast})\right]$.
Similarly we have
$1-F(x)=\left(  1-x^{\ast}\right)  \left[  1-\sigma(x^{\ast})x^{\ast}\right]$.
These allow us to write Eq. (\ref{M x' x general appendix}) as
\begin{align}
M(x^{\prime}|x)  &  =\binom{2N}{2Nx^{\prime}}\left\{  x^{\ast}\left[
1+\sigma(x^{\ast})(1-x^{\ast})\right]  \right\}  ^{2Nx^{\prime}}\left\{
\left(  1-x^{\ast}\right)  \left[  1-\sigma(x^{\ast})x^{\ast}\right]
\right\}  ^{2N(1-x^{\prime})}\nonumber\\
& \nonumber\\
&  =\binom{2N}{2Nx^{\prime}}\left(  x^{\ast}\right)  ^{2Nx^{\prime}}\left(
1-x^{\ast}\right)  ^{2N(1-x^{\prime})}\times\left[  1+\sigma(x^{\ast})(1-x^{\ast})\right]  ^{2Nx^{\prime}}\left[
1-\sigma(x^{\ast})x^{\ast}\right]  ^{2N(1-x^{\prime})}\nonumber\\
& \nonumber\\
&  \equiv M^{(0)}(x^{\prime}|x)\times e^{C(x^{\prime}|x)} \label{F1}
\end{align}
where
\begin{equation}
C(x^{\prime}|x)=2N\times\big[x^{\prime}\ln\big(1+\sigma(x^{\ast})\left(
1-x^{\ast}\right)  \big)+(1-x^{\prime})\ln\big(1-\sigma(x^{\ast})x^{\ast
}\big)\big]. \label{F2}
\end{equation}
Equation (\ref{F1}) represents an exact decomposition of the transition
matrix, $M(x^{\prime}|x)$, into the product of a transition matrix,
$M^{(0)}(x^{\prime}|x)$, that is independent of selection, and a factor
$e^{C(x^{\prime}|x)}$ which depends on selection, and is unity in the absence
of selection.

\subsection*{Factorisation of the transition matrix: \textit{n} alleles}

 For the case of $n$ alleles, the transition matrix can again be expressed as a product of 
 two factors,  one of which is independent of selection.

We begin with Eqs. (\ref{M multi}) and (\ref{fsel,i=xi*Gi}), which we
reproduce here for convenience:
\begin{equation}
M(\mathbf{x}^{\prime}|\mathbf{x})=\binom{2N}{2N\mathbf{x}^{\prime}}
\prod\nolimits_{i=1}^{n}\left[  F_{i}(\mathbf{x})\right]  ^{2Nx_{i}^{\prime} }
\label{M multi appendix}
\end{equation}
and
\begin{equation}
F_{i}(\mathbf{x})=x_{i}^{\ast}\times [1+G_{i}(\mathbf{x}^{\ast})]  \label{Fi appendix}
\end{equation}
in which
\begin{equation}
\mathbf{x}^{\ast}=\mathbf{Qx}. \label{x* mult appendix}
\end{equation}

In the absence of selection, $\mathbf{F}(\mathbf{x})$ reduces to
$\mathbf{x}^{\ast}$and $M(\mathbf{x}^{\prime}|\mathbf{x})$ reduces to
$M^{(0)}(\mathbf{x}^{\prime}|\mathbf{x})$, as given by
\begin{equation}
M^{(0)}(\mathbf{x}^{\prime}|\mathbf{x})=\binom{2N}{2N\mathbf{x}^{\prime}}
\prod\nolimits_{i=1}^{n}\left(  x_{i}^{\ast}\right)  ^{2Nx^{\prime}}.
\end{equation}

To establish a factorisation we note that Eq. (\ref{Fi appendix}) 
allows us to write Eq. (\ref{M multi appendix}) as
\begin{align}
M(\mathbf{x}^{\prime}|\mathbf{x})  &  =\binom{2N}{2N\mathbf{x}^{\prime}}
\prod\nolimits_{i=1}^{n}\left\{  x_{i}^{\ast}\left[  1+G_{i}(\mathbf{x}^{\ast
})\right]  \right\}  ^{2Nx_{i}^{\prime}}
=\binom{2N}{2N\mathbf{x}^{\prime}}\prod\nolimits_{i=1}^{n}\left(
x_{i}^{\ast}\right)  ^{2Nx_{i}^{\prime}}\times\prod\nolimits_{i=1}^{n}\left[
1+G_{i}(\mathbf{x}^{\ast})\right]  ^{2Nx_{i}^{\prime}}\nonumber\\
& \nonumber\\
&  =M^{(0)}(\mathbf{x}^{\prime}|\mathbf{x})\times e^{C(\mathbf{x}^{\prime
}|\mathbf{x})} \label{F1 mult}
\end{align}
where
\begin{equation}
M^{(0)}(\mathbf{x}^{\prime}|\mathbf{x})=\binom{2N}{2N\mathbf{x}^{\prime}}
\prod\nolimits_{i=1}^{n}\left(  x_{i}^{\ast}\right)  ^{2Nx_{i}^{\prime}}
\end{equation}
and
\begin{equation}
C(\mathbf{x}^{\prime}|\mathbf{x})=2N\sum\nolimits_{i=1}^{n}x_{i}^{\prime}
\ln\left(  1+G_{i}(\mathbf{x}^{\ast})\right)  . \label{F2 mult}
\end{equation}
Equation (\ref{F1 mult}) represents an exact decomposition of the transition
matrix, $M(\mathbf{x}^{\prime}|\mathbf{x})$, into the product of a transition matrix, $M^{(0)}(\mathbf{x}^{\prime}|\mathbf{x})$, that is independent of
selection, and a factor $e^{C(\mathbf{x}^{\prime}|\mathbf{x})} $ which depends
on selection, and is unity in the absence of selection.



\bibliography{WF}


\end{document}